\documentclass[runningheads, orivec]{llncs}
\usepackage[T1]{fontenc}
\usepackage{graphicx}
\usepackage{cite}
\usepackage{amsmath, amsfonts}
\usepackage{multirow}  
\usepackage{array}
\usepackage{booktabs} 
\usepackage[caption=false]{subfig}
\usepackage{tabularx} 
\usepackage{bm}
\usepackage{textcomp}
\usepackage{xcolor}
\usepackage{algorithm} 
\usepackage{algpseudocode}
\usepackage{xurl}
\usepackage{hyperref}

\newcommand{\showchanges}{0} 

\ifnum\showchanges=1
  \colorlet{revisioncolor}{blue}
\else
  \colorlet{revisioncolor}{black}
\fi

\begin{document}
\title{
Evaluating Grid Resilience in the Era of Ever-Increasing Data Centers}
%
%
\author{Yuhan Du\inst{1}\orcidID{0009-0004-4874-1235} \and Erika Ardiles-Cruze\inst{2} \and Javad Mohammadi\inst{1}\orcidID{0000-0003-0425-5302}}
%
%
\institute{University of Texas at Austin, Austin, TX 78705, USA \\
\email{\{yuhandu, javadm\}@utexas.edu} \and 
Air Force Research Laboratory, Rome, NY 13441, USA \\
\email{erika.ardiles-cruz@us.af.mil}}

\maketitle              

\makeatletter
\begingroup
\renewcommand{\thefootnote}{}
\def\Hy@footnote@currentHref{Hfootnote.AFRLclearance}
\footnotetext{This publication has been cleared under AFRL PA Number AFRL-2026-3175.}
\endgroup
\makeatother
%


\begin{abstract}
The rapid growth of artificial intelligence workloads is increasing the scale and concentration of data center demand, creating new concerns for power system resilience under disruptive events. 
This paper extends a validated multi-time-step DC optimal power flow framework to evaluate the impact of aggregated data center demand on contingency-induced unserved energy. 
Using an IEEE 30-bus system with flexible resources, we replace a conventional load at a contingency-exposed bus with an energy-matched constant data center load and examine two capacity-growth levels under generator derating, transmission line derating, and coupled derating. 
The results show that data center capacity growth substantially increases both system-level and data-center-bus unserved energy under transmission-constrained contingencies. 
Under coupled derating, the high-growth case increases total unserved energy from 3.203 MWh in the energy-matched case to 22.891 MWh. 
A supplementary energy-matched coincident-demand case further increases total unserved energy by 34.4\%, indicating that temporally concentrated data center demand can amplify resilience impacts even without increasing total energy consumption.

\end{abstract}

\section{Introduction} \label{s:introduction}
\subsection{Motivation}

Data centers are becoming an increasingly important source of electric demand as artificial intelligence (AI) and cloud computing services expand. 
In the United States, data center electricity use reached approximately 176 TWh in 2023, representing about 4.4\% of total electricity consumption, and is projected to reach 325-580 TWh by 2028, or approximately 6.7\%-12\% of total electricity consumption \cite{shehabi2024datacenter}. 
Unlike gradual growth in conventional demand, new data center loads can be large, concentrated, and sensitive to interruption. 
The North American Electric Reliability Corporation (NERC) has therefore identified emerging large loads, including data centers, as a growing reliability concern because their magnitude and operational characteristics can affect transmission planning, resource adequacy, and real-time system operation \cite{nerc2026largeloads}.

\begin{figure}[!t]
\centering
\color{revisioncolor}
\includegraphics[width=0.80\textwidth]{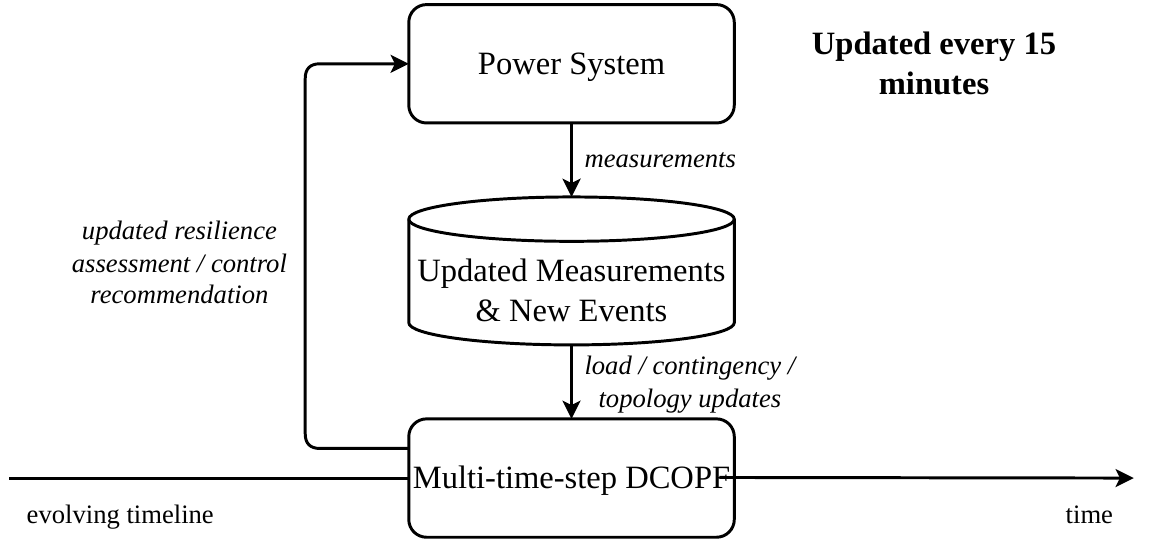}
\caption{Conceptual rolling-horizon DDDAS deployment of the proposed framework. Grid measurements and data center telemetry can update the system state at each 15-minute interval, after which the multi-time-step DCOPF can be re-solved to refresh resilience estimates and dispatch recommendations.}
\label{fig:workflow}
\vspace{-0.1in}
\color{black}
\end{figure}

In addition to their scale, AI-oriented data centers may introduce new temporal demand characteristics. 
A recent production data study of large AI training workloads reports synchronized power variations due to alternating computation and communication phases \cite{choukse2025power}. 
These fast variations are outside the temporal resolution of a dispatch model, but they motivate evaluating whether elevated data center demand overlapping with a grid disruption can further stress system operation. 
Recent data center modeling guidance also emphasizes that model fidelity should be selected based on the grid interaction being evaluated, ranging from detailed, fast-timescale representations to more tractable system-level abstractions \cite{talukdar2026modeling}. 
Accordingly, this work studies data centers as aggregated grid-facing loads over a dispatch horizon and includes a simplified coincident-demand sensitivity case.
\color{revisioncolor}
Figure \ref{fig:workflow} illustrates the intended DDDAS workflow: new measurements and contingency information update the model state on a rolling basis, and the multi-time-step DCOPF is re-solved periodically.
\color{black}

\subsection{Literature Review}

Early studies on data center demand response demonstrated that computing workloads and local generation can be coordinated to reduce coincident peak demand and electricity expenditure \cite{liu2013datacenter}. 
More recent studies have extended this line of work toward grid-interactive data center operation. 
Wan et al. incorporated Internet data center demand response into electricity network transition planning and evaluated its effect using an IEEE 30-bus system \cite{wan2023internet}. 
Zhang et al. developed a joint data center and load aggregator optimization framework for electricity demand response \cite{zhang2024aggregator}, while Liu et al. considered computing workload shifting within the energy management of a data center microgrid \cite{liu2024microgrid}. 
These studies indicate that data center demand can be represented as an operationally relevant grid-side resource rather than only as a passive load.

The growing interaction between data centers and power systems has also motivated broader flexibility assessments. 
Recent reviews evaluate data centers as a source of flexibility for power systems and assess the evidence for grid-service provision from data center demand \cite{takci2025flexibility,crozier2025potential}. 
Zhang et al. further developed a coordinated scheduling strategy to mitigate power-grid impacts from proactive data center workload shifts \cite{zhang2025mitigating}. 
These studies primarily examine flexibility provision, energy management, or grid-interactive scheduling.

Power system resilience, however, concerns performance during disruptive conditions rather than only operational flexibility under routine conditions. 
Panteli and Mancarella introduced a conceptual framework for power system resilience under high-impact events \cite{panteli2015grid} and subsequently developed a quantitative assessment model for critical electrical infrastructure \cite{panteli2017modeling}. 
More recently, Golan et al. reviewed resilience quantification in evolving power grids and highlighted the importance of critical function, temporal resolution, and disruption benchmarks in evaluating system performance \cite{golan2026dimensions}. 
In our prior work, a multi-time-step DC optimal power flow (DCOPF) model was used to quantify unserved load under generator and transmission derating while coordinating flexible resources in the IEEE 30-bus system \cite{du2025resilience}. 
Building on that validated framework, the present study examines how the magnitude and temporal concentration of aggregated data center demand affect contingency-induced unserved energy. 
Unlike previous data center flexibility studies, the analysis uses an energy-matched comparison to separate load-profile effects from capacity-growth effects and introduces a dispatch-scale coincident-demand case to evaluate temporal concentration during disruptions.

\subsection{Contributions}
The main contributions of this work include:
\begin{itemize}
\item Extending a validated multi-time-step DCOPF resilience framework with an aggregated data center load representation at a contingency-exposed node.
\item Developing an energy-matched comparison to evaluate the effects of data center load profiles and capacity growth under generator derating, transmission line derating, and their combined effects.
\item Introducing a dispatch-scale coincident-demand sensitivity case to quantify how temporally concentrated data center demand can affect system-level and data-center-bus unserved energy.
\end{itemize}

\section{Methodology}  \label{s:methodology}
\subsection{Multi-Time-Step Dispatch Framework}

We extend the multi-time-step DCOPF resilience model developed in \cite{du2025resilience}. 
At each time step $t$, the model determines available generation, battery charging and discharging, power flows, and emergency unserved demand while considering projected load and disruption conditions over a finite decision horizon. 
Let $P^{G}_{i,t}$ denote generation at bus $i$, $P^{dis}_{i,t}$ and $P^{ch}_{i,t}$ denote battery discharge and charge, and $S_{i,t}$ denote unserved demand. 
The optimization objective is represented as

\begin{equation}
\min \sum_{t}\left(C^{G}_{t}+C^{B}_{t}+c^{S}\sum_{i}S_{i,t}\right),
\label{eq:objective}
\end{equation}
where $C^{G}_{t}$ and $C^{B}_{t}$ represent generation and battery operation costs, and $c^{S}$ assigns a high cost to unserved demand.

The nodal power balance and principal operating limits are written as
\begin{align}
P^{G}_{i,t}+P^{dis}_{i,t}-P^{ch}_{i,t}+S_{i,t}-L_{i,t}
    &= \sum_{j} F_{ij,t}, \label{eq:balance}\\
-\overline{F}_{ij,t} \leq F_{ij,t} &\leq \overline{F}_{ij,t}, \label{eq:line}\\
0 \leq P^{G}_{i,t} &\leq \overline{P}^{G}_{i,t}, \label{eq:gen}\\
E_{i,t} &= E_{i,t-1}+\eta P^{ch}_{i,t}\Delta t
          -\frac{P^{dis}_{i,t}\Delta t}{\eta}. \label{eq:soc}
\end{align}
Battery energy and charging/discharging powers are further restricted by their operating bounds. 
In the reproduced baseline implementation, emergency unserved demand is allowed at buses with positive modeled demand using a common interruption cost. 
We retain this treatment for comparability and separately report unserved energy at the selected data center bus.

\subsection{Aggregated Data Center Load Model}

The data center is represented as an aggregated grid-facing demand at a selected bus $d$. 
Let $L^{O}_{d,t}$ denote the original time-varying load profile at this bus. 
To separate temporal load-profile effects from increased demand, the energy-matched constant data center demand (DC) is defined as
\begin{equation}
L^{DC}_{d,t} = \overline{L}^{DC}_{d} =
\frac{1}{N_T}\sum_{t=1}^{N_T}L^{O}_{d,t}
\label{eq:dc-matched}
\end{equation}
Capacity-growth cases are then represented by
\begin{equation}
L^{DC,\beta}_{d,t}=\beta \overline{L}^{DC}_{d},
\qquad \beta \in \{1.5,2.0\}.
\label{eq:dc-growth}
\end{equation}

We additionally consider one dispatch-scale coincident-demand sensitivity case for the high-growth data center profile. 
Let $L^{HG}_{d}=2\overline{L}^{DC}_{d}$ denote the constant high-growth data center demand, and let $\mathcal{T}_{D}$ denote the disruption-active time steps. 
The swing proxy is defined as
\begin{equation}
L^{SW}_{d,t} =
\begin{cases}
(1+\alpha)L^{HG}_{d}, 
    & t \in \mathcal{T}_{D}, \\[2pt]
\left(1-\alpha
\displaystyle\frac{|\mathcal{T}_{D}|}
{N_T-|\mathcal{T}_{D}|}\right)L^{HG}_{d},
    & t \notin \mathcal{T}_{D}.
\end{cases}
\label{eq:swing}
\end{equation}
where $\alpha=0.20$. 
This construction preserves total data center energy demand over the horizon while concentrating higher demand during disruption intervals. 
It is a dispatch-scale sensitivity representation rather than a reproduction of fast facility-level power dynamics.

\subsection{Resilience Metrics}

Resilience is evaluated using total unserved energy and data-center-bus unserved energy (UE):
\begin{equation}
{UE}^{sys}=\Delta t\sum_{t}\sum_{i}S_{i,t},
\qquad
{UE}^{DC}=\Delta t\sum_{t}S_{d,t}.
\label{eq:ue}
\end{equation}
For the data center bus, we also report the served-energy ratio:
\begin{equation}
R^{DC}=1-\frac{UE^{DC}}
{\Delta t\sum_{t}L_{d,t}}.
\label{eq:served-ratio}
\end{equation}

\color{revisioncolor}
\section{Relation to the DDDAS Paradigm}  \label{s:dddas}
Dynamic Data Driven Applications Systems (DDDAS) couple a physical system and a computational model through dynamic measurements and model-guided sensing or control. 
Figure \ref{fig:workflow} illustrates a conceptual rolling-horizon deployment of the proposed framework. 
At each update interval, grid measurements, data center telemetry, and contingency information could update the system state and forecast trajectories, after which the multi-time-step DCOPF could be re-solved to refresh dispatch and unserved-energy estimates.

\begin{table}[!ht]
\color{revisioncolor}
\centering
\small
\caption{Mapping of the proposed framework to the DDDAS feedback-control loop.}
\label{tab}
\setlength{\tabcolsep}{4pt}
\renewcommand{\arraystretch}{1.03}
\begin{tabularx}{\textwidth}{@{}>{\raggedright\arraybackslash}p{0.27\textwidth}>{\raggedright\arraybackslash}X@{}}
\toprule
DDDAS function & Role in the proposed framework \\
\midrule
Physical system and dynamic data &
Grid measurements, data center telemetry, and contingency updates define the evolving operating state. \\
\addlinespace
Data augmentation and simulation &
Measurements update system states and forecast trajectories, and the multi-time-step DCOPF estimates dispatch and unserved energy. \\
\addlinespace
Feedback and instrumentation &
Updated system-level and data-center-bus risk can prioritize sensing and inform candidate storage, workload, local-generation, or demand-response actions. \\
\bottomrule
\end{tabularx}
\color{black}
\end{table}
\vspace{-0.1in}

The proposed architecture therefore supports the data-augmentation and model-driven feedback functions of the DDDAS paradigm. The numerical experiments in this paper evaluate the underlying time-coupled optimization using prescribed time-indexed load and contingency trajectories. Real-time telemetry integration, repeated rolling-horizon execution, and automated actuation remain implementation steps for closing the full feedback-control loop.

\color{black}

\section{Case Study Setup}  \label{s:case_study}
The proposed data center cases are evaluated using the IEEE 30-bus system with the flexible-resource settings from the reproduced baseline model \cite{du2025resilience}. 
The system includes batteries at buses 2, 8, 15, and 27 and solar generation at buses 14 and 30.
Each simulation uses a 15-minute resolution and a 36-step decision horizon, corresponding to 9 hours. 
This horizon is selected because the reproduced baseline and prior results show no further reduction in unserved demand beyond 36 steps for the coupled-derating case.

Bus 7 is selected as the data center location. 
It has an original active demand of 22.8 MW, no existing battery, and is located near the generation and transmission assets affected by the modeled contingencies. 
It therefore represents a contingency-exposed stress test location rather than a general siting recommendation. 
Over the 9-hour horizon, the original Bus 7 profile has a total demand of 236.543 MWh, an average demand of 26.283 MW, and a peak demand of 30.723 MW. 
Accordingly, the energy-matched constant data center demand is set to 26.283 MW. 
The two capacity-growth cases are set to 39.424 MW and 52.565 MW, corresponding to 1.5$\times$ and 2$\times$ the energy-matched demand, respectively.
\color{revisioncolor}
The 1.5$\times$ and 2$\times$ growth levels and the 20\% coincident-demand increase are sensitivity settings rather than facility-specific forecasts; disruption timing and deratings are retained from the baseline model for comparability.
\color{black}

Three contingency scenarios are included from the baseline study. 
In the generator derating case, the generation capacities at buses 1 and 2 are reduced by 90\%. 
In the line derating case, lines 1-2, 1-3, 2-4, 3-4, 2-5, and 2-6 are reduced to 5\% of their nominal flow limits. 
The coupled case applies both deratings simultaneously. 
Disruptions occur during three 15-minute intervals within the simulated peak-period window. Finally, the coincident-demand sensitivity case is applied only to the high-growth data center profile under coupled derating. 
It increases demand by 20\% during the disruption-active intervals while preserving the same total 9-hour demand of 473.086 MWh as the constant high-growth case.
We simulate this study in a Python 3.6 environment operating on a MacBook Pro (Apple M3 Max, 2023).

\section{Results and Discussion}  \label{s:results}
\subsection{Baseline Reproduction and Energy-Matched Results}

The validated baseline model was first reproduced using the 36-step, 9-hour decision horizon. 
Under coupled derating, the reproduced load-shedding sum is 24.762 MW-periods, equivalent to 6.190 MWh at the 15-minute resolution, matching the previously reported result. 
Since longer decision horizons did not further reduce unserved load in the prior study, the 9-hour horizon is used for the following data center experiments.

Table \ref{tab:main-dc-results} reports total unserved energy and unserved energy at the data center bus under the three contingency scenarios. 
The original profile and DC-matched cases have identical Bus-7 demand of 236.543 MWh over the 9-hour horizon. 
The DC-matched case uses a constant demand of 26.283 MW, while the original profile reaches a peak demand of 30.723 MW. 
Consequently, the constant energy-matched representation reduces peak-period exposure in this evaluation window and produces lower unserved energy than the original time-varying profile.

\begin{table}[t]
\centering
\small
\setlength{\tabcolsep}{2.5pt}
\caption{Unserved energy under different load and contingency scenarios. Each entry reports ``total unserved energy (UE) / data center (DC) bus UE'' in MWh.}
\label{tab:main-dc-results}
\begin{tabular}{lccc}
\toprule
Load Case & Gen. & Line & Coupled \\
\midrule
Original profile   & 0.941 / 0.000  & 6.065 / 2.963   & 6.190 / 3.101 \\
DC matched         & 0.000 / 0.000  & 3.079 / 0.000   & 3.203 / 0.113 \\
DC growth 1        & 4.053 / 0.000  & 12.921 / 9.819  & 13.047 / 9.957 \\
DC growth 2        & 9.060 / 0.000  & 22.766 / 19.665 & 22.891 / 19.803 \\
\bottomrule
\end{tabular}
\end{table}

\subsection{Capacity-Growth and Coincident-Demand Effects}

Data center capacity growth substantially increases unserved energy, particularly when transmission delivery is constrained. 
Under line derating, total unserved energy increases from 3.079 MWh in the DC-matched case to 22.766 MWh in the high-growth case, while data center bus unserved energy increases from 0 to 19.665 MWh. 
Under coupled derating, total unserved energy similarly increases from 3.203 to 22.891 MWh, with 19.803 MWh occurring at the data center bus. 
These results indicate that, for the contingency-exposed location examined here, increased data center capacity primarily aggravates local service interruption under line-constrained disruptions.

To examine temporal demand concentration, the constant high-growth profile is compared with an energy-matched coincident-demand case under coupled derating. 
Both profiles require 473.086 MWh at Bus 7 over the 9-hour horizon; however, the coincident-demand case increases data center demand during the disruption-active intervals and offsets this by reducing demand in the remaining intervals.

\begin{figure}[H]
\centering
\includegraphics[width=0.52\textwidth]{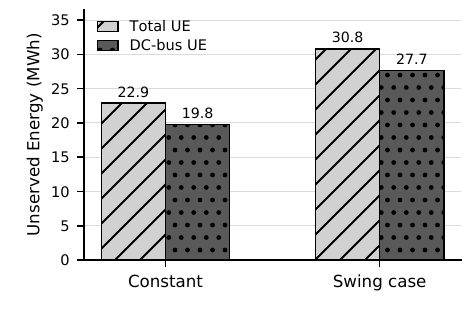}
\vspace{-0.1in}
\caption{Effect of disruption-coincident demand under the high-growth coupled-derating case; both profiles have identical 9-hour energy demand.}
\label{fig:swing-comparison}
\end{figure}

As shown in Fig. \ref{fig:swing-comparison}, disruption-coincident high demand increases total unserved energy from 22.891 to 30.777 MWh, corresponding to a 34.4\% increase. 
Data center bus unserved energy increases from 19.803 to 27.688 MWh, or 39.8\%, while the data center served-energy ratio decreases from 95.81\% to 94.15\%. 
This sensitivity case indicates that temporally concentrated data center demand can materially amplify contingency impacts even when total energy demand is unchanged. 
The swing case is represented at the 15-minute dispatch timescale and does not reproduce second-scale power dynamics of AI training facilities.

\color{revisioncolor}
\subsection{Operational Implications and Scope}

Grid-interactive storage, local generation, and workload-aware demand response are potential mitigation options, and in a DDDAS deployment their setpoints could be updated as grid and facility data arrive. 
The reported values are specific to the IEEE 30-bus system, Bus 7, and the selected deratings. The framework can also represent weather-driven events through time-varying generation and line availability, or cyber events through delayed, missing, or corrupted measurements. 
Broader siting studies, a wider range of workload-growth profiles, and contingency-severity analyses remain future work.
\color{black}

\section{Conclusion}  \label{s:conclusion}

This paper applies a validated multi-time-step DCOPF resilience framework to aggregated data center demand under generator and transmission contingencies.
In the studied setting, an energy-matched constant data center profile produces lower unserved energy than the original peak-period conventional profile, whereas capacity growth substantially increases system-level and data-center-bus unserved energy under transmission constrained disruptions.
An energy-matched coincident-demand case further increases total unserved energy by 34.4\% under coupled derating, highlighting the importance of temporal demand concentration.
Future work will consider higher-fidelity data center models, broader siting and contingency-severity studies, and mitigation through grid-interactive storage and controllable computing demand.

\bibliographystyle{splncs04}
\bibliography{conference.bib}

\end{document}